\documentstyle[12pt,german]{article}
\begin{document}
\begin{Large}
\begin{center}
{\bf A more comprehensive formulation of evolutionary equations}\\
\end{center}
\end{Large}
\begin{large}
\begin{center}
{Ingrid Hartmann}\\
\end{center}
\end{large}
\footnote{Ingrid Hartmann, Querstrasse 4a, 18442 Buschenhagen, Germany}
\begin{abstract}
As mathematical model for the evolutionary equations of species the 
masterequation is choiced. Two formulations will be demonstrated  
to include the changes of parameters into the masterequation - 
that is, on the one hand, the formation of a second masterequation
for the development of parameters, and, on the other hand, the use
of the Wigner-distribution to describe the development of parameters.
Moreover, the Wigner-distribution is used to describe morphic fields
and involved in the theorie of selforganization.
\end{abstract}
\setcounter{tocdepth}{2}
\tableofcontents
\section{Introduction}
In the course of the past few years, in the frame of the theory of
selforganization, attempts have been made by a great number of authors
\cite{Eigen,E.F,P.N,H.S1} to describe the development (evolution)
of biological systems. In this way, the competition of species,
which, in each case, have a certain amount of characteristic features, 
could be simulated and the dynamic behaviour for certain attempts
in the growing and decay rates could be determined. However, these
attempts lead to stagnation as soon as a stationary behaviour is reached.
The reason is that - in the author's opinion - the change of the parameters of
species (for instance of the growing rates), which could change the behaviour
once achieved, was not included in the equations in an adequate way.
Attempts have been made \cite{E.F,H.S1,E.S1,E.S2} to change these
features of species with the help of mutation terms via chance, so that it
turns out to be possible basically (in principle) to reach another section 
of the parameter room that shows another dynamic behaviour. In this way, a 
more or less diffuse walk through the parameter room is possible.  
In some papers we have tried to solve these problems \cite{H.S2,H.S3}
assuming that the development of the parameters during the evolution process
happens in a way that the function, which is characteristic for the
particular species, becomes optimal. However, these attempts have had the 
disadvantage that the discovering of this efficiency function - provided
that it actually exists - is subject to a certain arbitrariness. On the 
other hand, this attempt contains the additional assumption that the 
development of the parameters is not inherently contained in the evolutionary
equations. In the present paper, the author is going to demonstrate another
way that allows to include changes of the parameters dirctly into the 
evolutionary equations of species.
\section{Behaviour under constant parameters}
The behaviour under constant parameters will be demonstrated at the following
example (a detailed description is given in \cite{E.F}). The total number
of sorts amounts to $ N = N_{1} + N_{2} $. The sort $ N_{1} $ increases with  
growing rate $ E_{1} $ and the sort $ N_{2} $ with the growing rate
$ E_{2} $. The elementary stochastic process is the growing of one sort at
the expense of the other sort.
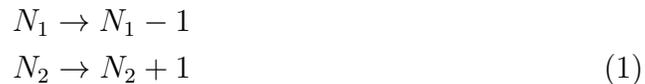
\begin{eqnarray}\label{gleichung1}
N_{1} \rightarrow N_{1} - 1\nonumber\\
N_{2} \rightarrow N_{2} + 1
\end{eqnarray}
It is:
\begin{eqnarray}\label{gleichung1000}
N_{1} + N_{2} & = & const..
\end{eqnarray}
$ a_{ij} $ be the rate of the spontaneous transition from j to i.
For the corresponding transition probabilities valids:
\begin{eqnarray}\label{gleichung1001}
W^+(N_{2}) & = & E_{2}N_{1}(N_{2}/N) + a_{21}\nonumber\\
W^-(N_{2}) & = & E_{1}N_{2}(N_{1}/N) + a_{12}
\end{eqnarray}
The masterequation can then be written for the probability $ P(N_{2},t) $:
\begin{eqnarray}\label{gleichung2}
\frac{\partial P(N_{2},t)}{\partial t} & = & [a_{21} + \frac{E_{2}}{N}
(N_{2} - 1)(N - N_{2} + 1)]P(N_{2} -1,t)\nonumber\\
& & + [a_{12} + \frac{E_{1}}{N}(N_{2} + 1)(N - N_{2} - 1)]P(N_{2} + 1,t)\nonumber\\
& & - [a_{12} + a_{21} + \frac{1}{N}(E_{1} + E_{2})N_{2}(N - N_{2})]P(N_{2},t)
\end{eqnarray}
The stationary solution is shown in \cite{E.F}.\\
In the stationary case with stochastic description a probability distribution
$ P^{eq}(N_{2}) \neq 0 $ for $ N_{2} \neq 0 $ is achieved, that means
that the two sorts can coexist with each other in a certain manner,
but in the deterministic case the sort with the greatest growing rate wins.
(However, this field of problems shall not be dealt with in the present
connection.) Now, how can the change of the features $ a_{ij}, E_{i} $ be 
included in order to meet the conditions of the gradual changes of the
parameters? In the next sections two proposals will be made to that matter.
\section{Behaviour under time-varying parameters}
On the basis of equation \ref{gleichung2} the following attempt shall be made:
\begin{eqnarray}\label{gleichung3}
\frac{\partial P(N_{2},t)}{\partial t} & = & [\langle a_{21} \rangle
+ \frac{\langle E_{2} \rangle}{N}(N_{2} - 1)(N - N_{2} + 1)]
P(N_{2} - 1,t)\nonumber\\
& & + [\langle a_{12} \rangle + \frac{\langle E_{1} \rangle}{N}(N_{2} + 1)
(N - N_{2} -1)]P(N_{2} + 1,t)\nonumber\\
& & - [\langle a_{12} \rangle + \langle a_{21} \rangle + \frac{1}{N}
(\langle E_{1} \rangle + \langle E_{2} \rangle)N_{2}(N - N_{2})]
P(N_{2},t).
\end{eqnarray}
The parameters $ a_{ij}, E_{i} $ ara time-dependent and have a specific
eigentime $ \tau _{ij}, \tau_{i} $, according to which they are 
developing.\\ 
As simplefied consideration the following attempt shall be made:
\begin{center}
$ a_{12} $ develops in the course of time to $ a_{(12)'} $,\\
$ a_{21} $ develops in the course of time to $ a_{(21)'} $,\\
$ E_{1}  $ develops in the course of time to $ E_{1'} $,\\
$ E_{2}  $ develops in the course of time to $ E_{2'} $.
\end{center}
It valids:
\begin{eqnarray}\label{gleichung4}
a_{12} + a_{(12)'} & = &  a^{(1)}_{m} = const.\nonumber\\
a_{21} + a_{(21)'} & = &  a^{(2)}_{m} = const.\nonumber\\
E_{1} + E_{1'} & = & E^{(1)} = const.\nonumber\\
E_{2} + E_{2'} & = & E^{(2)} = const. .
\end{eqnarray}
The description of the temporal development be demonstrated by
the example of $ E_{1}(t) $. $ E_{1}(t) $ obey the following equation:\\
\begin{eqnarray}\label{gleichung5}
W^+(E_{1}) & = & d_{1'1}E_{1'} = d_{1'1}(E^{(1)} - E_{1})\nonumber\\
W^-(E_{1}) & = & d_{11'}E_{1} .
\end{eqnarray}
The masterequation 
\begin{eqnarray}\label{gleichung6}
\frac{\partial P_{E^{(1)}}(E_{1},t)}{\partial t} & = & W^+(E_{1} - 1,t)
P(E_{1} - 1,t) + W^-(E_{1} + 1,t)P(E_{1} + 1,t)\nonumber\\
& & - [W^+(E_{1},t) + W^-(E_{1},t)]P(E_{1},t) . 
\end{eqnarray}
then reads as follows:
\begin{eqnarray}\label{gleichung7}
\frac{\partial P_{E^{(1)}}(E_{1},t)}{\partial t} & = & d_{1'1}
[E^{(1)} - (E_{1} - 1)]
P(E_{1} - 1,t) + d_{11'}(E_{1} + 1)P(E_{1} + 1,t)\nonumber\\
& & - [d_{1'1}(E^{(1)} -E_{1}) + d_{11'}E_{1}]P(E_{1},t) . 
\end{eqnarray}
The stationary solution says:\\
\begin{eqnarray}\label{gleichung8}
P_{E^{(1)}}^{eq}(E_{1}) & = & {E^{(1)} \choose E_{1}} \left[\frac{d_{1'1}}
{d_{1'1} + d_{11'}}\right]^{E_{1}}\left[\frac{d_{11'}}{d_{1'1} + d_{11'}}
\right]^{E^{(1)} - E_{1}}\nonumber\\
\langle E_{1}(t \to \infty) \rangle & = & \left( \frac{d_{1'1}}
{d_{1'1} + d_{11'}} \right) E^{(1)} = pE^{(1)}\nonumber\\
\langle E_{1}(t \to \infty) \rangle & = & \left( \frac{d_{11'}}
{d_{1'1} + d_{11'}} \right) E^{(1)} = (1 - p)E^{(1)}\nonumber\\
mit\nonumber\\
p & = & \frac{d_{1'1}}{d_{1'1} + d_{11'}} . 
\end{eqnarray}
Consequently, the time-dependent solution is:
\begin{eqnarray}\label{gleichung9}
\langle E_{1}(t) \rangle & = & pE^{(1)} + (E_{0} - pE^{(1)})
\exp(-(d_{1'1} + d_{11'})t)\nonumber\\
\langle E_{1'}(t) \rangle & = & E^{(1)} - \langle E_{1}(t) \rangle = 
E^{(1)}(1 - p)\nonumber\\
& & + (pE^{(1)} - E_{0})\exp(- (d_{1'1} + d_{11'})t)\nonumber\\
mit\nonumber\\
E_{0} - starting value.
\end{eqnarray}  
$ P(t) $ is given in \cite{M.S}.\\
Analogous to this the equations for $ E_{2}(t) $ can be formulated.
In the case of strong temporal separation of the development of the 
species from the development of the features follows equation   
\ref{gleichung3} with the stationary solutions:
\begin{eqnarray}\label{gleichung10}
\langle E_{1}(\infty) \rangle & = & pE^{(1)}\nonumber\\
\langle E_{2}(\infty) \rangle & = & qE^{(2)}\nonumber\\
with\nonumber\\
q & = & \frac{d_{2'2}}{d_{2'2} + d_{22'}} .
\end{eqnarray}
By analogy the considerations for $ a_{12} $ and $ a_{21} $ can be 
continued. In the case of a not so strong temporal separation, that is if the  
time-development of $ E_{j} $ has to be considered as well, follows from 
equation \ref{gleichung3} and \ref{gleichung9} :\\
\begin{eqnarray}\label{gleichung11}
\frac{\partial P(N_{2},t)}{\partial t} & = & \nonumber\\
\langle a_{12} \rangle P(N_{2} - 1,t)\nonumber\\
+ (E^{(2)}q + (E_{02} - qE^{(2)})\exp(-(d_{2'2} + d_{22'})t))
\frac{(N_{2} -1)}{N}(N - N_{2} + 1)P(N_{2} - 1,t)\nonumber\\
+ \langle a_{21} \rangle P(N_{2} + 1,t)\nonumber\\
+ (E^{(1)}p + (E_{01} - pE^{(1)})\exp(-(d_{1'1} + d_{11'})t))
\frac{(N_{2} + 1)}{N}(N - N_{2} - 1)P(N_{2} + 1,t)\nonumber\\
- [\langle a_{12} \rangle + \langle a_{21} \rangle ]P(N_{2},t)\nonumber\\
- \frac{1}{N} (E^{(1)}p + E^{(2)}q)N_{2}(N - N_{2})P(N_{2},t)\nonumber\\
- \frac{1}{N} (E_{02} - qE^{(2)})\exp(d_{2'2} + d_{22'})t)
N_{2}(N - N_{2})P(N_{2},t)\nonumber\\
- \frac{1}{N} (E_{01} - pE^{(1)})\exp(d_{1'1} + d_{11'})t)
N_{2}(N - N_{2})P(N_{2},t)
\end{eqnarray}
Thereby, the solution of this equation will become extremely complicated.
For the present, it may remain here as an attempt proposal. In an analogous
way, the other terms can be dealt with as well. For space reasons,
this question will not be considered in the present paper.
\section{Behaviour under time-varying parameters with the help of the
description by the Wigner-distribution}
A further possibility to include the development of the parameters
in the masterequation offers the following attempt with the help of the
Wigner-distribution.\\
$ E_{i} $ be the feature of a species which in the course of the time t 
can be changed.\\  
It has turned out to be necessary to introduce - in addition to the
''normal'' time t - characteristic times for the particular species.
So let $ E_{i} $ be reverse proprotional to a characteristic eigentime  
$ \tau_{i} $ of the species:
\begin{eqnarray}\label{gleichung12}
E_{i} & \sim  & \frac{1}{\tau_{i}}
\end{eqnarray}
and so it valids:
\begin{eqnarray}\label{gleichung13}
E_{i} & \sim & \omega_{i}
\end{eqnarray}
with $ \omega_{i} $ the characteristic frequency of a species. (The principle
of the method, for simplification, shall be demonstrated at one frequency
only. It can be extended to a characteristic frequency spectrum). 
(Considerations about different times can be found in \cite{Prig,H.S2,H.S3}.)\\
Now the attempt shall be made:
\begin{eqnarray}\label{gleichung14}
s^{(i)} & = & \sum_{j=1}^n \cos(E_{j}^{(i)}t)\exp(-\alpha_{j}(t - t_{j})^2),
\end{eqnarray}
by which it shall be described that the species have the feature 
$ E_{j}^{(i)} $ at a certain time $ t_{j} $ hat. Furthermore, the attempt
shows that the sort can potentially have all frequencies $ \omega_{j}^{(i)} $ 
however actually it happens at certain times $ t_{j} $. In this way, each sort
is characterised by a tupel $ (E_{j}^{(i)},t_{j},\alpha_{j}) $ 
or $ (\omega_{j}^{(i)},t_{j},\alpha_{j}) $ respectively, and so it represents
an energy-time-unit.\\  
From quantum mechanics the Wigner-distribution is known:
\begin{eqnarray}\label{gleichung15}
W(q,p) & = & \int^{\infty}_{-\infty}\exp(-\frac{j2 \pi x}{h})\phi(q+\frac{x}{2})
\phi^*(q-\frac{x}{2})\,dx 
\end{eqnarray}
($ \phi $ - wave function, q - possition, p - momentum, h - Planck's constant).\\
In electrical engineering it is applied as follows:
\begin{eqnarray}\label{gleichung16}
W(\omega,t) & = & \frac{1}{2\pi}\int^{\infty}_{-\infty}
\exp(-j\tau\omega)f(t+\frac{\tau}{2})f^*(t-\frac{\tau}{2})\,d\tau 
\end{eqnarray}
($ f(t) $ - signal, $ \omega $ - frequency).\\
In this case, it serves for the analysis of time-varying spectrums and a
distribution over $ \omega $ and t.\\
In our case, if we consider $ s_{i}(t) $ according to equation \ref{gleichung14}
as ''wave-function'' of the species, it means that
\begin{eqnarray}\label{gleichung17}
W(\omega,t) & = & \frac{1}{2\pi}\int^{\infty}_{-\infty}
\exp(-jE\tau)s(t+\frac{\tau}{2})s^*(t-\frac{\tau}{2})\,d\omega . 
\end{eqnarray}
demonstrates the probability of the features of the species. The mean
property $ \langle E_{j}^{(i)} \rangle $ can be obtained by formation of the
moments of the Wigner-distrbution \cite{Me}, and it can be included in this
form in the masterequation \ref{gleichung3}.\\
In the following, these considerations will be demonstrated in detail for the 
case $ n = 2 $. From equation \ref{gleichung14} and equation \ref{gleichung17} 
follows then:
\begin{eqnarray}\label{gleichung18}
W^{(1)} & = & W_{11}^{(1)} + W_{12}^{(1)} + W_{21}^{(1)} + W_{22}^{(1)} .
\end{eqnarray}
Consequently, the Wigner-distrbution, according to equation \ref{gleichung17},
is composed of two autocorrelation terms $ W_{ii}^{(1)} $ and two
crosscorrelation terms $ W_{ij}^{(1)} $ which result from terms with equal
or different indices respectively.\\
The calculation gives the following terms:
\begin{eqnarray}\label{gleichung19}
\lefteqn{W_{11}^{(1)}  = }\nonumber\\
& & \frac{1}{4} \sqrt{\frac{\pi 4}{\alpha}}
\exp[-\alpha (t - t_{1})^2]\nonumber\\
& & [2\cos(2E_{1}t)\exp(-\frac{E^2}{\alpha}) + \exp(-\frac{(E + E_{1})^2}
{\alpha}) + \exp(-\frac{(E - E_{1})^2}{\alpha})]\nonumber\\
\lefteqn{W_{22}^{(1)}  = }\nonumber\\
& & \frac{1}{4} \sqrt{\frac{\pi 4}{\alpha}}
\exp[-\alpha (t - t_{2})^2]\nonumber\\
& & [2\cos(2E_{2}t)\exp(-\frac{E^2}{\alpha}) + \exp(-\frac{(E + E_{2})^2}
{\alpha}) + \exp(-\frac{(E - E_{2})^2}{\alpha})]\nonumber\\
\lefteqn{W_{Kreuz} =  W_{12}^{(1)} + W_{21}^{(1)}  =}  \nonumber\\
& & \frac{1}{4} \sqrt{\frac{\pi 2}{\alpha}} \exp(-\alpha(t - t_{1})^2) 
\exp(-\alpha(t - t_{2})^2)\nonumber\\
& & \left(2\cos[(E_{1} + E_{2})t]
\left[\exp(-\frac{[E + \frac{(E_{2} - E_{1})}{2}]^2}{2\alpha})
+ \exp(-\frac{[E - \frac{(E_{2} - E_{1})}{2}]^2}{2\alpha})
\right]\right.\nonumber\\
& & + \left.2\cos[(E_{2} - E_{1})t]
\left[\exp(-\frac{[E + \frac{(E_{2} + E_{1})}{2}]^2}{2\alpha})
+ \exp(-\frac{[E - \frac{(E_{2} + E_{1})}{2}]^2}{2\alpha})
\right]\right) 
\end{eqnarray} 
(For simplification it was calculated with
$ \alpha_{1} = \alpha_{2} = \alpha $.)
From the calculation of the local moment of the  Wigner-distribution\\
\begin{eqnarray}\label{gleichung20}
\langle E(t) \rangle & = & \frac{1}{2\pi} \int^{\infty}_{0}EW(t,E)\,dE/p_{s}
\nonumber\\
with\nonumber\\
p_{s} & = & \frac{1}{2\pi} \int^{\infty}_{\infty}W(t.E)\,dE
\end{eqnarray}
results in special, with $ p_{s} $ being the momentary energy of s:
\begin{eqnarray}\label{gleichung21}
\langle E_{11}^{(1)}(t) \rangle & = & \frac{1}{4\pi} \sqrt{\frac{\pi}{\alpha}}
\exp(-\alpha(t - t_{1})^2) \nonumber\\
& & \left[2\cos(2E_{1}t)\frac{\alpha}{2} + \frac{\alpha}{2} + E_{1}
\frac{\sqrt{\pi \alpha}}{2} \right]/p_{s} \nonumber\\
\langle E_{22}^{(1)}(t) \rangle & = & \frac{1}{4\pi} \sqrt{\frac{\pi}{\alpha}}
\exp(-\alpha(t - t_{2})^2) \nonumber\\
& & \left[2\cos(2E_{2}t)\frac{\alpha}{2} + \frac{\alpha}{2} + E_{2}
\frac{\sqrt{\pi \alpha}}{2} \right]/p_{s} \nonumber\\
\langle E_{12}^{(1)}(t) \rangle & = & \frac{1}{8\pi} \sqrt{\frac{2\pi}{\alpha}}
\exp(-\alpha(t - t_{1})^2)\exp(-\alpha(t - t_{2})^2) \nonumber\\
& & \left(2\cos((E_{1} + E_{2})t)\left[\alpha + (\frac{E_{2} - E_{1}}{2})
\frac{\sqrt{\pi \alpha 2}}{2} \right]\right. \nonumber\\
& & \left.+ 2\cos((E_{2} - E_{1})t)\left[\alpha + (\frac{E_{2} + E_{1}}{2})
\frac{\sqrt{\pi \alpha 2}}{2} \right] \right)/p_{s} \nonumber\\
\langle E^{(1)}(t) \rangle & = & \langle E_{11}^{(1)}(t)\rangle
+ \langle E_{22}^{(1)}(t) \rangle
+ \langle E_{12}^{(1)}(t) \rangle  
\end{eqnarray}
These expressions for the mean values can be included into the masterequation
\ref{gleichung3}. In this way, we can get a more extensive description.  
(The other parameters can be dealt with in a analogous way.) So the
development of the parameters can directly be included into the 
masterequation and a walk from pik to pik in the probabilty distribution   
$ P(N_{2},t) $ can happen because of the fact that at certain times
certain feature values are adopted so that again new relations with
other sorts are in another way formed.
\section{Conclusions}
The author considers the Wigner-distribution is an adequate description
for species since it can describe the time-development and,
at the same time, the frequency-development of the species (or the
development of the characteristic times of species respectively).
The author, in a further publication, will come back to these problems.
In the present paper, there is only given a short outline. The
Wigner-distribution is considered as probability distribution on the
time behaviour of species and, therefore, represents an ''information
field''.\\
To inform means to bring in shape. According to the present attempt,
certain values $ E_{i} $ are taken at certain times $ t_{i} $. Potential
features $ E_{i} $ are made real features $ E_{i} $, which means they are 
brought in shape. In this respect, the Wigner-distribution can be considered
as information distribution for species. This description, in the author's 
opinion, corresponds with the description given by R.Sheldrake \cite{Sh1,Sh2},
who describes the develpoment of features with the help of morphic fields.
The morphic field as probability - and information field \cite{Sh2} is 
described in the present attempt of the Wigner-distribution.\\
Furthermore, it can be demonstrated \cite{Puneu} that the Wigner-distribution
included as probability distribution in the relation 
\begin{eqnarray}\label{gleichung100}
S & = & k\ln(W)
\end{eqnarray}
produces additional terms which in the theorie of selforganization 
contribute to structure formation.\\
In this way, the term \bf{information} \rm takes up a central place
in the theory of selforganization \cite{Puneu} \\
\begin{center}
- as mental potential that,\\
  at a certain time, \\
is brought in shape - . 
\end{center}
The author thinks that the present considerations could be one step 
forward on the way to integrate the term \bf {consciousness} \rm
in the ''world of physics'' - an idea which recently has repeatedly
been considered by a number of famous physicists \cite{Capra,Wigner,Bohm,Sh3}.        

\end{document}